\documentclass[preprint,twocolumn,10pt,a4paper,byrevtex,aip,superscriptaddress,showpacs]{revtex4-1}
\usepackage{natbib}
\usepackage{graphicx}
\usepackage{amsfonts}
\usepackage{amsmath}
\usepackage{amssymb}
\usepackage{color}

\begin{document}

\title{Detection of spin torque magnetization dynamics through low frequency noise}

\author{Juan Pedro Cascales*}
\email{(*)juanpedro.cascales@uam.es}
\affiliation{Dpto. Fisica Materia Condensada C3, Instituto Nicolas Cabrera (INC), Condensed Matter Physics Institute (IFIMAC), Universidad Autonoma de Madrid, Madrid 28049, Spain}

\author{David Herranz}
\affiliation{Dpto. Fisica Materia Condensada C3, Instituto Nicolas Cabrera (INC),
Condensed Matter Physics Institute (IFIMAC), Universidad Autonoma de Madrid, Madrid 28049, Spain}

\author{Ursula Ebels}
\affiliation{SPINTEC, UMR 8191, CEA/CNRS/UJF \& G-INP, INAC, 38054 Grenoble Cedex, France}

\author{Jordan A. Katine}
\affiliation{Hitachi Global Storage Technologies, San Jose, California 95135, USA}

\author{Farkhad G. Aliev\textdaggerdbl}
\email{(\textdaggerdbl) farkhad.aliev@uam.es}
\affiliation{Dpto. Fisica Materia Condensada C3, Instituto Nicolas Cabrera (INC), Condensed Matter Physics Institute (IFIMAC), Universidad Autonoma de Madrid, Madrid 28049, Spain}

\begin{abstract}

We present a comparative study of high frequency dynamics and low frequency noise in elliptical 
magnetic tunnel junctions with lateral dimensions under 100 nm presenting current-switching phenomena. The analysis of the high frequency 
oscillation modes with respect to the current reveals the onset of a steady-state precession regime for negative bias currents above $J=10^7 A/cm^2$, when the magnetic field 
is applied along the easy axis of magnetization. By the study of low frequency noise for the same samples, we demonstrate the direct link between
changes in the oscillation 
modes with the applied current and the normalised low frequency (1/f) noise as a function of the bias current.
These findings prove that low frequency noise studies could be a simple and powerful technique 
to investigate spin-torque based magnetization dynamics.

\end{abstract}

\maketitle


Oscillations known as a ferromagnetic resonance arise from the precessional motion of the magnetization 
of a ferromagnetic material (FM) when an external magnetic field is applied in the presence of microwave 
pump field perpendicular to it. Since magnetic tunnel junctions (MTJs) are composed of ferromagnetic electrodes,
and exhibit the tunneling magnetoresistance effect \cite{Julliere,Moodera1995,Miyazaki1995}, 
it is possible to detect magnetization dynamics through the measurement of frequency dependent votage noise power 
(typically up to a few tens of GHz) in DC biased MTJs (see the review \cite{RalphStiles}). 
In the regime where the current density applied to the MTJ is low, the
resulting damped oscillatory modes are due to the external applied magnetic field and 
thermal fluctuations, referred to as thermal FMR (T-FMR). 
This effect is typically observed for applied current densities
below $J\simeq10^{7} A/cm^{2}$\cite{RalphStiles}. The effective damping can be cancelled
altogether by the spin torque \cite{SlonLLG} (ST) from a d.c., spin-polarized current at some critical 
value of the current density $J_C$. This results in an auto-oscillation of the magnetization 
which is often referred to as a steady state precession. The ability to switch the magnetic state of MTJs with only current could
pave the way for new, smaller and faster data storage devices. 
Using MTJs with lateral sizes under 100 nm would increase the storage density
of devices, reduce their power consumption and contribute to the development of current controlled 
microwave sources.

The transition from the T-FMR regime to an in-plane, steady state precession (SSP)
can be identified from a sudden decrease in the frequency and linewidth $\Delta f$ 
on the applied current \cite{Kim2008,Tiberkevich2007,Mistral2006,Petit2007,HouseSTT}.
The T-FMR/SSP transition in fact presents two regimes with critical currents $J_C$, $J_{C}^{*}$, where for $|J_C|<|J|<|J_C^{*}|$ the
system presents an intermittent steady state (stable for a few $ns$) 
with linewidths in the hundreds of MHz, which becomes stable for several $\mu$s when $|J_C^*|<|J|$
and presents linewidths an order of magnitude lower\cite{HousePRL}. 

Our work shows that MTJ samples with a magnetic field along the easy axis present spin-torque effects
so the coercive field of the free electrode is shifted to lower values. Most importantly, the influence
of changes with the current in the high frequency (HF) oscillation modes has also been observed in low frequency (LF) noise measurements. 
Previous studies indicate that the low frequency tail in the HF noise power could be affected by the transition from a damped oscillation to a 
steady state precession \cite{Devolder2009,Thadani2008}. 
To our best knowledge, no direct evidence of such a relation has been obtained so far through a systematic study. 
The LF measurements presented here may constitute a better quantification of 
the stochastic hopping at the transition between the T-FMR to SSP 
in the kHz range. Also, this realization could be widely useful, as measurements in the kHz range 
are technically simpler than for high frequency signals (MHz-GHz).


\begin{figure}[h!]
\begin{center}
\includegraphics[width=0.47\textwidth]{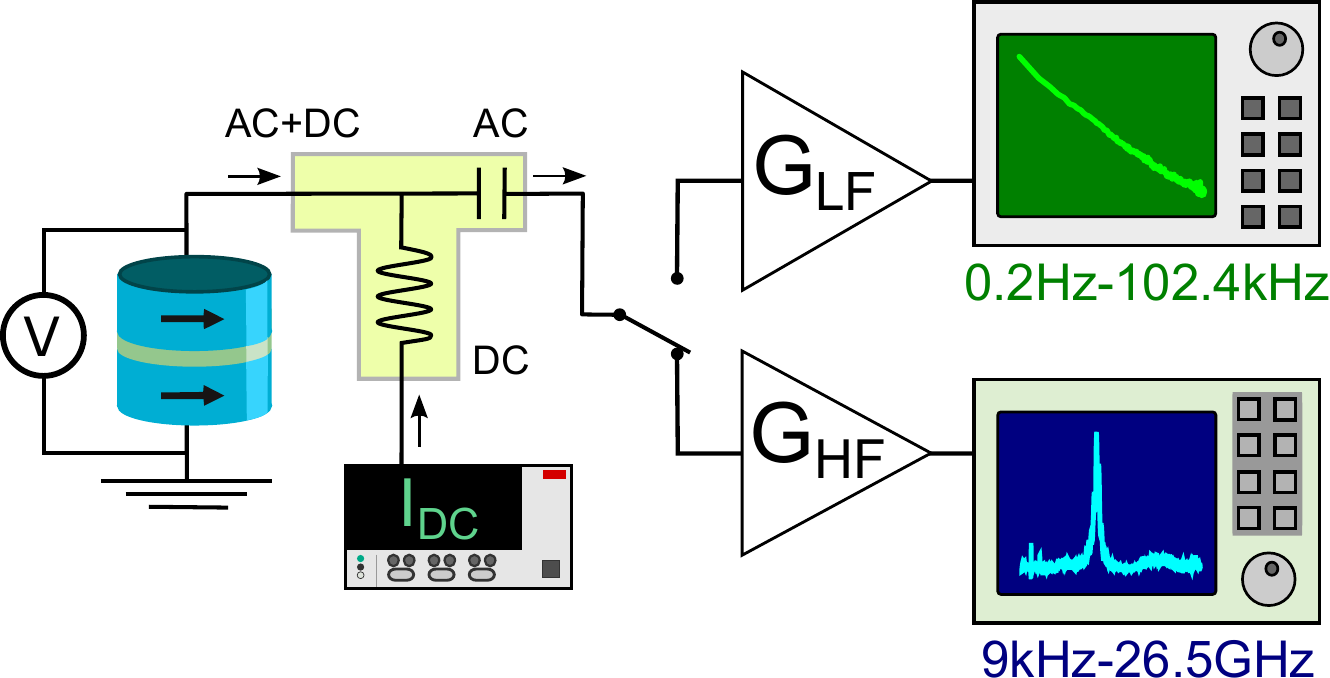} 
\end{center}
\caption{Diagram of the experimental set-up where either a low (kHz range) or high frequency (GHz range) noise measurement scheme
can be selected.}
\label{fig:fig1}
\end{figure}

The multilayer MTJ nanopillars have the following structure: 
IrMn(6.1)/ CoFeB(1.8)/ Ru/ CoFe(2)/ MgO(0.9nm)/ CoFe(0.5)/CoFeB(3.4) 
where the numbers indicate the thickness of the layers in nm. The pinned layer consists
of two FM layers which are antiferromagnetically coupled through a thin ruthenium layer.
The lower FM layer is exchange-coupled to an antiferromagnetic IrMn layer. 
The MgO barrier is deposited by sputtering and the free 
layer consists of a bi-layer of CoFe/CoFeB.  The measured nanopillar devices have
elliptical cross-sections of different sizes, with the minor and major axes ranging 
from 40$\times$80 to 65$\times$130 (in nm), and a nominal RA product of 1.5 $\Omega\times\mu m^2$. 
The easy axis (EA) direction is parallel to the pinned layer's magnetization coinciding with the major axis of the ellipse, 
while the in-plane hard axis (HA) is perpendicular (but still in-plane) to the EA. The devices are 
embedded in impedance matched RF coplanar waveguides for electrical contacting using 
special RF probes. Figure \ref{fig:fig1} shows a diagram of the experimental set-up. 
The samples are biased by a d.c. current which is 
input through the LF port of a bias tee, and the voltage across the device is measured by a nanovoltmeter.
The voltage signal from the sample goes through the mixed port of the bias tee and is input into 
either LF or HF system through the HF port of the bias tee. The LF system was previously described in Ref. \cite{CascalesAPL2013}.
Regarding the HF measurement, the voltage fluctuations out of the HF port are
amplified by a Miteq AVG6 amplifier, and then input into an Agilent Technologies EXA signal analyzer (bandwidth 9 kHz-26.5 GHz).
Details of the calibration of the HF setup may be found in Ref. \cite{tesisjp}.

Previous measurements on devices of this kind\cite{HouseSTT,HousePRL,ThesisHousse}
have shown that these MTJs fall into two different groups:
samples with high resistance and tunneling magnetoresistance (TMR) ratios around 90\% (labeled HTMR)
and samples with low resistance and TMR ratios around 30-60\% (labeled LTMR). 
The authors report that repetead high-current 
measurements on HTMR devices may gradually turn them into LTMR devices. These LTMR devices
seem to be stable against high-current measurements, and the authors speculate 
that the difference between sample types
could be due to localized reductions in the tunneling barrier\cite{HouseSTT}.
Indeed, the statistics of our MTJs reveal a mean TMR value of around 60\% 
and an average resistance-area (RA) product of 1.8 $\Omega\times\mu m^2$ (in agreement
with the nominal value of 1.5 $\Omega\times\mu m^2$).
MTJs with ultra-low RA are important for practical devices. For junctions with low MgO thicknesses (and low RA products, correspondingly), there exists a correlation between the TMR
ratio and the MgO thickness\cite{Yuasa2004}. A decrease in TMR is experienced when the
thickness of the MgO barrier is reduced, due to local inhomogeneities (or pseudo-pinholes) in the MgO barrier. 
Obtaining MTJs with low RA products and high TMR ratios, for which homogeneous barriers are needed, 
is a real engineering challenge. The STT effect requires high current densities, and since MgO barriers 
only withstand a certain amount of voltage, having a low RA allows high currents to flow
through the MTJ without causing the breakdown of the barrier\cite{ThesisAlvarez}. 
We shall focus below on LTMR MTJs, as they are best
candidates to study magnetization dynamics and spin torque effects.

\begin{figure}[htb!]
   \begin{center}
   \includegraphics[width=0.47\textwidth]{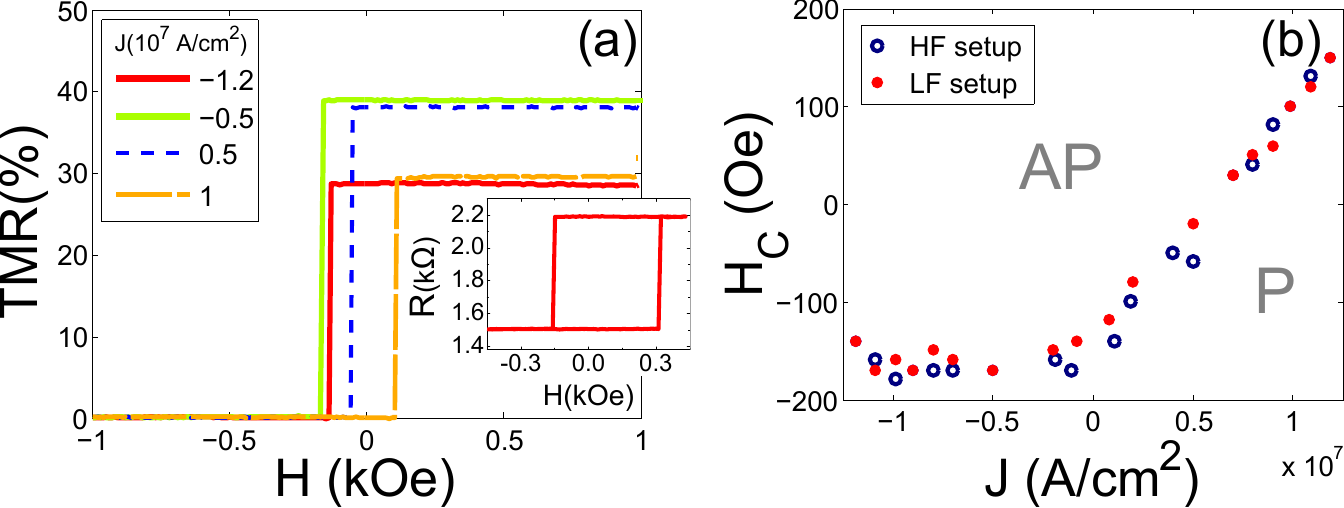}
  \caption{40x80nm$^2$ MTJ biased along the easy axis direction. (a) TMR curves at different applied currents. 
  The coercive field of the MTJ changes with increasing positive current. The inset shows a full TMR cycle at (b) 
  Change in the coercive field with the applied current from HF and LF noise measurements.}s\label{fig:fig2}
\end{center}
  \end{figure}

Figure \ref{fig:fig2} shows that when the external field is directed along the easy axis, one observes a step-like transition in resistance, 
from a high resistance state at positive fields (antiparallel or AP state)
to a lower resistance state at negative fields (parallel or P state). In our measurements, positive voltage means that electrons 
flow from the pinned to the free electrode, promoting the P state. Negative voltage favors the AP state. 
The high and low frequency noise measurements shown in this work were carried out in the same 40x80nm$^2$ sample, 
although qualitatively similar results were obtained for several others.
Fig. \ref{fig:fig2}(a) shows changes in the coercive field with the applied current, obtained from resistive transitions measured by sweeping the field 
positive to negative values. A full TMR cycle taken at low bias (2 mV) is shown in the inset of Fig. \ref{fig:fig2}(a).
As could be seen in Fig. \ref{fig:fig2}(b), the coercive
field $H_C$ is stable for negative currents (the AP state is favored) while shifts to lower values 
(favoring the P state) for the positive currents. Note that the results obtained from both high and
low frequency set-ups have been plotted in Fig. \ref{fig:fig2}(b).

\begin{figure}[h!]
\begin{center}
\includegraphics[width=0.47\textwidth]{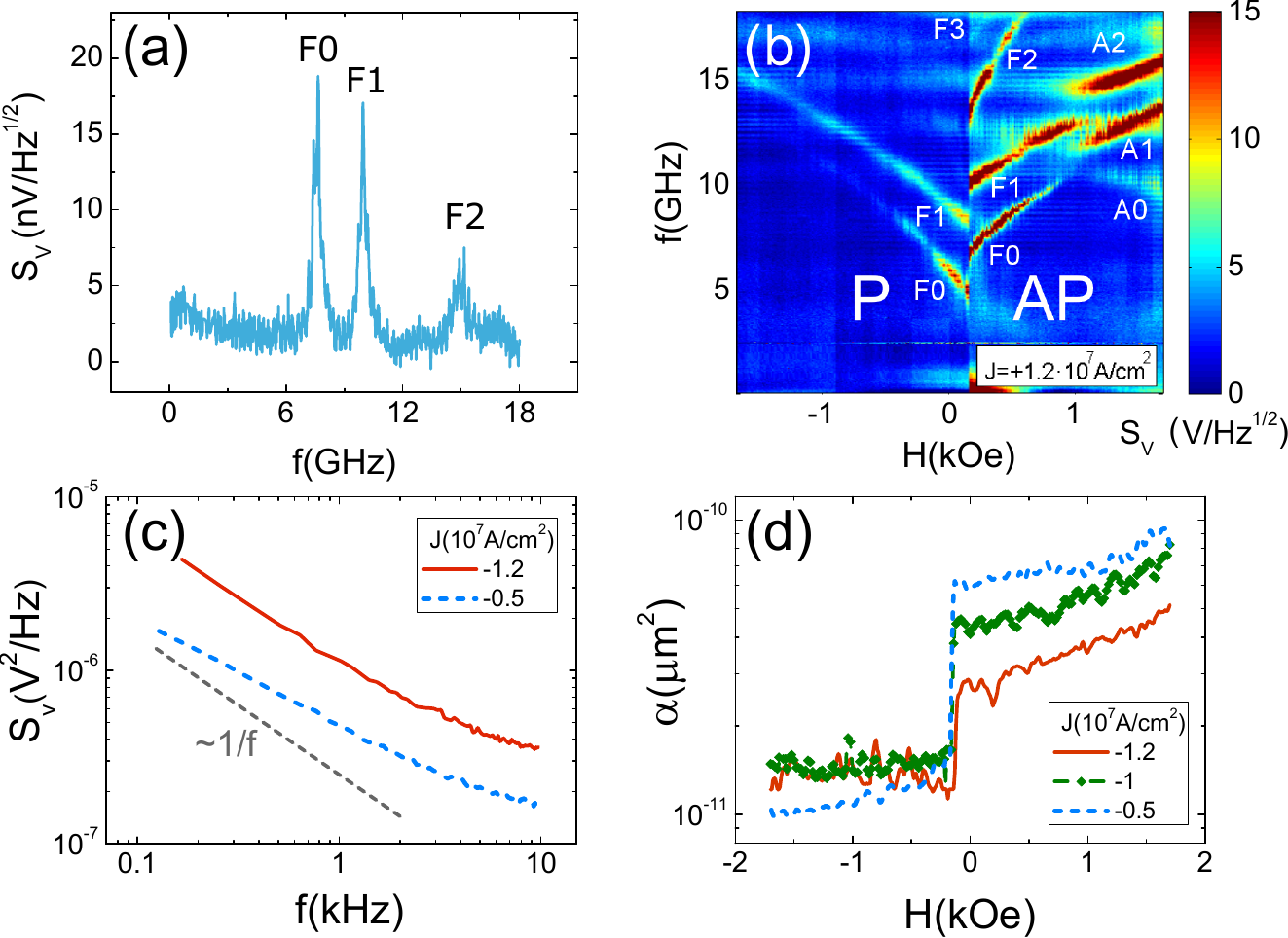}
\end{center}
  \caption{(a) High frequency spectrum in the P state (H=-350 Oe) for $J=-1.2\cdot 10^7 A/cm^2$ of
a 40x80nm elliptical MTJ. The spectrum reveals 3 different free layer oscillation modes, F0, F1 and F2. (b) Surface plots at $J=+1.2\cdot10^7$ A/cm$^2$ constructed 
  with the high frequency spectra, presenting several oscillation modes of the FM layers. 
  (c) Noise power spectra in the P state (H=-350 Oe) where the dashed line depicts a $\sim 1/f^{0.8}$ dependence and (d) dependence of the  Hooge parameter on the external field for several current density values.}\label{fig:fig3}
\end{figure}

A typical high frequency noise spectrum presents resonance peaks 
centered around a frequency $f_{res}$, with linewidths $\Delta f$. 
An example of such a spectrum is shown in Fig. \ref{fig:fig3}(a).
We have studied the evolution of these resonance modes with
both an external magnetic field and a d.c. current $I$. The resonance peaks may
be characterized by their $f_{res}$, $\Delta f$ and output
power $P_{out}$ of the microwave emission.
We have constructed surface plots (see Fig. \ref{fig:fig3}(b)) at constant current, with the high frequency spectra taken at different
applied external fields, so the evolution of the modes with the current can be detected. 
For positive currents, the P state is stabilized and the AP
state is destabilized, and vice versa for negative currents. This is reflected in the fact that 
the modes observed have higher amplitudes in the AP state for positive currents, and in the P state
for negative currents. 

\begin{figure}[!h]
\begin{center}
\includegraphics[width=0.47\textwidth]{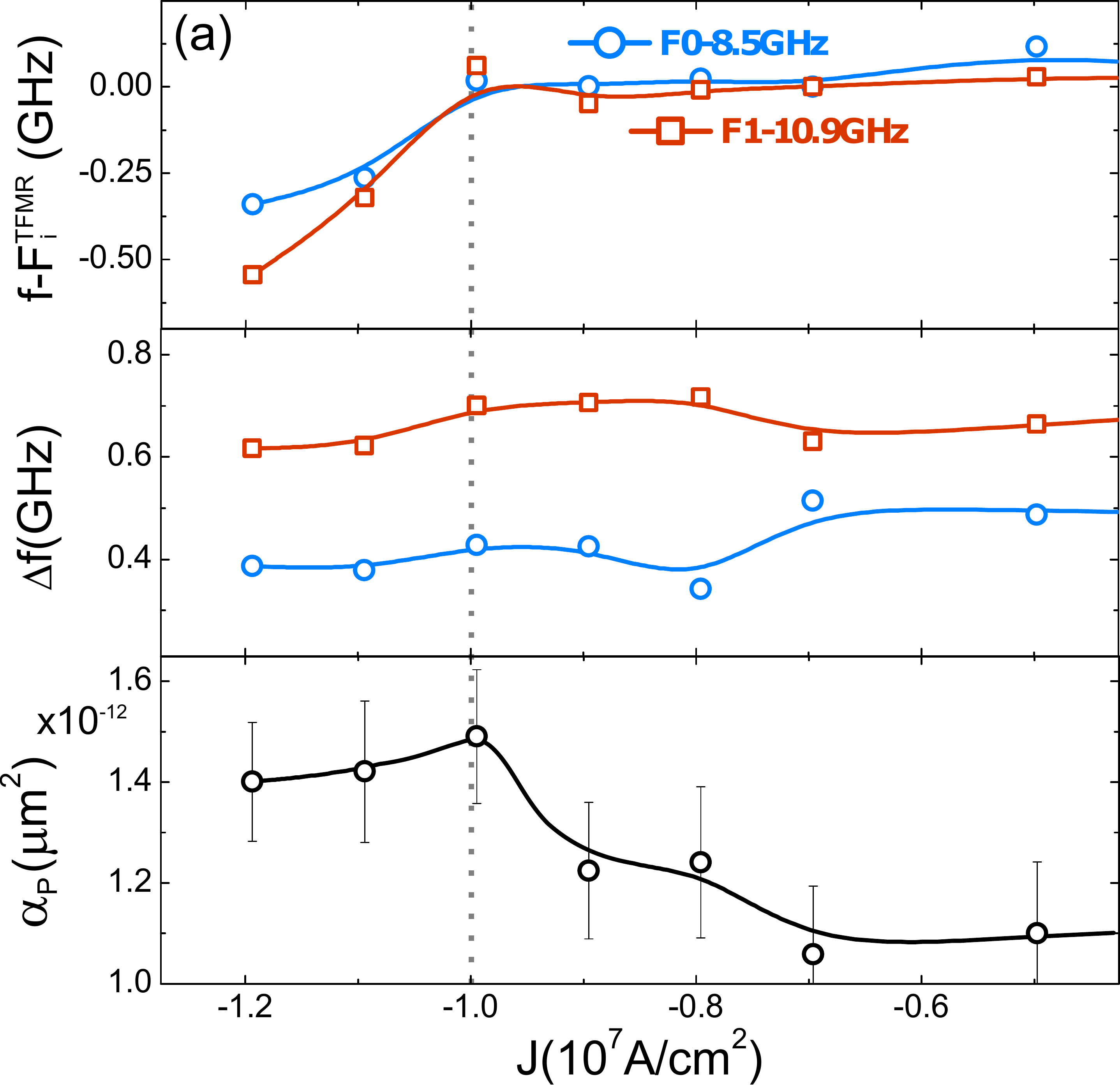} 
\end{center}
\caption{High and low frequency noise measurements in the P state (-350 Oe)  of a 40x80nm$^2$ MTJ. 
Dependence of the F0 and F1 modes' (a) frequency shift with respect to the modes' frequency in the TFMR regime, (b) linewidth and (c) average Hooge parameter in the P
state, $\alpha_P$, as a function of the current density.}
\label{fig:fig4}
\end{figure}

Six clear modes are detected, which come from oscillations
of the free layer (labeled F0, F1, ...) and the SAF structure (A0, A1 and A3).
The free layer modes are V-shaped, while the modes not showing a minimum at low fields
correspond to SAF modes\cite{Helmer2010}. The SAF modes should present a minimum at the high field required 
for the spin-flop of the SAF, but our applied fields are not high enough.
The F0 mode typically corresponds to excitations localized near the edges of the layer\cite{Helmer2010}.
The F2 mode only appears in the state which is excited (P or AP), depending
on the polarity of the current. A possible fourth free layer mode, F3, is labeled, although
it appears very tenuous and is only present for $J=+1.2\cdot 10^7 A/cm^2$. Several
other samples revealed similar oscillation modes. 
We have carried out an analysis of the F0 and F1 modes (as they have the highest amplitudes)
in the manner discussed in Ref.\cite{HouseSTT}. 
The analysis of these results reveals that for negative currrents, 
a decrease in frequency is observed for the first and second modes in the
P state, starting at $J\sim - 10^7 A/cm^2$. Under the same conditions, the AP state data presents the same dependence as the P state,
but the features are not as clear. The change of the oscillation frequency with respect to the TFMR regime and the 
linewidth of the F0 and F1 modes for the P state (H=-350 Oe) is shown in Fig. \ref{fig:fig4}(a) and (b).
As can be seen, the decrease in frequency of the F0 and F1 modes resembles the transition from T-FMR to SSP
reported on similar devices in Ref.\cite{HouseSTT}.
The minimum linewidth $\Delta f$ obtained for the F0 mode is around 400 MHz, which agrees with
what was previously observed for the intermittent steady state in Ref.\cite{HousePRL}.
The microwave power was found to monotonically increase with the current and is not shown for brevity.
Therefore we conclude that our highest negative current takes our sample to an intermittent SSP regime\cite{HousePRL}. 
For positive currents (not shown) 
such a transition is not found, and the emitted microwave power and low frequency noise monotonically increase with 
the applied current.

Low frequency noise measurements were also carried out in the same sample, using the same
current and magnetic field values. The low frequency spectra (see Fig. \ref{fig:fig3}(c)) may be described by $S_{V}(f)= \frac{\alpha V^2}{A f^{\beta}}$ where
$V$ is the applied voltage, $A$ the sample area, $\alpha$ is the normalized 1/f noise or Hooge parameter and $0.8<\beta <1.5$. 
The extraction of the 1/f parameters was carried out by carrying out a linear fit to $log(S_V)=log(\alpha V^2/A)-\beta log(f)$ between
$0<f<10kHz$.
The Hooge parameter remains somewhat constant for each magnetic state (P or AP) (see Fig. \ref{fig:fig3}(d). 
The analysis of the 1/f noise data as a function of the current density reveals the signature of effects
observed in the HF results. As is shown in Fig. \ref{fig:fig4}(c), the Hooge parameter $\alpha$ 
in the P state, $\alpha_P$, monotonically increases with
the applied current for current densities substantially below $\pm 10^7 A/cm^2$. This trend is observed regardless of the
field range within the P state chosen for the average.
Then, only for negative currents, the normalized noise reaches a maximum and starts decreasing
at around $J=-10^7 A/cm^2$. 

1/f noise in spin torque oscillators has been tied to hopping of the oscillation modes, where
each random hopping event leads to a jump in the phase of the oscillator\cite{Eklund2014,Sharma2014}.
As the current is increased towards the critical value for an intermittent SSP\cite{HousePRL} ($J_C=-10^7 A/cm^2)$,
these phase jumps increase in number, which is reflected in 1/f results. When the intermittent SSP
is reached and the current continues increasing, the hopping events gradually become less frequent
until eventually reaching the purely SSP for a second critical current\cite{HousePRL} $J_C^*$ (not reached in our experiment).
So by comparing our low frequency results with the high frequency data, 
we can ascertain that we are detecting signs of spin-torque
related phenomena in low frequency 1/f measurements. 
Further, our low frequency results seem to be more sensible to ST-driven effects than high frequency measurements,
since the 1/f noise begins increasing for lower current densities than the ones needed to observe any change 
in the oscillation modes (TFMR-SSP transition).

We remark that qualitatively different low frequency
noise was observed in HTMR junctions with spin torque effects being suppressed in the current range under study. These MTJs, expected to
have a more uniform barrier revealed a decrease in the Hooge factor with an increasing applied 
bias, similarly to what was previously observed for Fe/MgO/Fe MTJs with 2-3 nm thick MgO barriers\cite{Gokce2006,Aliev2007,Almeida2008}.


\emph{In summary}, current switching effects have been observed in low TMR nanopillar MTJs of 
sizes under 100nm \cite{HouseSTT,HousePRL,ThesisHousse} if the MTJs are biased along the easy axis, where an AP/P 
switch is favored for positive currents. The analysis of the high frequency oscillation modes with respect to the 
bias current reveals the onset of a steady-state precession regime for negative currents,
when the field is applied along the easy axis. A comparison of this analysis with 1/f noise as a 
function of the current shows that the changes in magnetization dynamics in the GHz range are 
reflected in the low frequency noise. The beginning of the transition to the steady state
regime appears as a maximum in the normalized 1/f noise (Hooge parameter).
The obtained results should help to define the ``current window
range'' for the potential application of nm sized magnetic
tunnel junctions by using LF noise measurement techniques.


\begin{thebibliography}{24}
\expandafter\ifx\csname natexlab\endcsname\relax\def\natexlab#1{#1}\fi
\expandafter\ifx\csname bibnamefont\endcsname\relax
  \def\bibnamefont#1{#1}\fi
\expandafter\ifx\csname bibfnamefont\endcsname\relax
  \def\bibfnamefont#1{#1}\fi
\expandafter\ifx\csname citenamefont\endcsname\relax
  \def\citenamefont#1{#1}\fi
\expandafter\ifx\csname url\endcsname\relax
  \def\url#1{\texttt{#1}}\fi
\expandafter\ifx\csname urlprefix\endcsname\relax\def\urlprefix{URL }\fi
\providecommand{\bibinfo}[2]{#2}
\providecommand{\eprint}[2][]{\url{#2}}

\bibitem[{\citenamefont{Julliere}(1975)}]{Julliere}
\bibinfo{author}{\bibfnamefont{M.}~\bibnamefont{Julliere}},
  \bibinfo{journal}{Physics Letters A} \textbf{\bibinfo{volume}{54}},
  \bibinfo{pages}{225} (\bibinfo{year}{1975}).

\bibitem[{\citenamefont{Moodera et~al.}(1995)\citenamefont{Moodera, Kinder,
  Wong, and Meservey}}]{Moodera1995}
\bibinfo{author}{\bibfnamefont{J.~S.} \bibnamefont{Moodera}},
  \bibinfo{author}{\bibfnamefont{L.~R.} \bibnamefont{Kinder}},
  \bibinfo{author}{\bibfnamefont{T.~M.} \bibnamefont{Wong}}, \bibnamefont{and}
  \bibinfo{author}{\bibfnamefont{R.}~\bibnamefont{Meservey}},
  \bibinfo{journal}{Phys. Rev. Lett.} \textbf{\bibinfo{volume}{74}},
  \bibinfo{pages}{3273} (\bibinfo{year}{1995}).

\bibitem[{\citenamefont{Miyazaki and Tezuka}(1995)}]{Miyazaki1995}
\bibinfo{author}{\bibfnamefont{T.}~\bibnamefont{Miyazaki}} \bibnamefont{and}
  \bibinfo{author}{\bibfnamefont{N.}~\bibnamefont{Tezuka}},
  \bibinfo{journal}{Journal of Magnetism and Magnetic Materials}
  \textbf{\bibinfo{volume}{139}}, \bibinfo{pages}{0} (\bibinfo{year}{1995}).

\bibitem[{\citenamefont{Ralph and Stiles}(2008)}]{RalphStiles}
\bibinfo{author}{\bibfnamefont{D.}~\bibnamefont{Ralph}} \bibnamefont{and}
  \bibinfo{author}{\bibfnamefont{M.}~\bibnamefont{Stiles}},
  \bibinfo{journal}{Journal of Magnetism and Magnetic Materials}
  \textbf{\bibinfo{volume}{320}}, \bibinfo{pages}{1190} (\bibinfo{year}{2008}).

\bibitem[{\citenamefont{Slonczewski}(1996)}]{SlonLLG}
\bibinfo{author}{\bibfnamefont{J.}~\bibnamefont{Slonczewski}},
  \bibinfo{journal}{Journal of Magnetism and Magnetic Materials}
  \textbf{\bibinfo{volume}{159}}, \bibinfo{pages}{0} (\bibinfo{year}{1996}).

\bibitem[{\citenamefont{Kim et~al.}(2008)\citenamefont{Kim, Mistral, Chappert,
  Tiberkevich, and Slavin}}]{Kim2008}
\bibinfo{author}{\bibfnamefont{J.-V.} \bibnamefont{Kim}},
  \bibinfo{author}{\bibfnamefont{Q.}~\bibnamefont{Mistral}},
  \bibinfo{author}{\bibfnamefont{C.}~\bibnamefont{Chappert}},
  \bibinfo{author}{\bibfnamefont{V.~S.} \bibnamefont{Tiberkevich}},
  \bibnamefont{and} \bibinfo{author}{\bibfnamefont{A.~N.}
  \bibnamefont{Slavin}}, \bibinfo{journal}{Phys. Rev. Lett.}
  \textbf{\bibinfo{volume}{100}}, \bibinfo{pages}{167201}
  (\bibinfo{year}{2008}).

\bibitem[{\citenamefont{Tiberkevich et~al.}(2007)\citenamefont{Tiberkevich,
  Slavin, and Kim}}]{Tiberkevich2007}
\bibinfo{author}{\bibfnamefont{V.}~\bibnamefont{Tiberkevich}},
  \bibinfo{author}{\bibfnamefont{A.}~\bibnamefont{Slavin}}, \bibnamefont{and}
  \bibinfo{author}{\bibfnamefont{J.-V.} \bibnamefont{Kim}},
  \bibinfo{journal}{Applied Physics Letters} \textbf{\bibinfo{volume}{91}},
  \bibinfo{eid}{192506} (\bibinfo{year}{2007}).

\bibitem[{\citenamefont{Mistral et~al.}(2006)\citenamefont{Mistral, Kim,
  Devolder, Crozat, Chappert, Katine, Carey, and Ito}}]{Mistral2006}
\bibinfo{author}{\bibfnamefont{Q.}~\bibnamefont{Mistral}},
  \bibinfo{author}{\bibfnamefont{J.-V.} \bibnamefont{Kim}},
  \bibinfo{author}{\bibfnamefont{T.}~\bibnamefont{Devolder}},
  \bibinfo{author}{\bibfnamefont{P.}~\bibnamefont{Crozat}},
  \bibinfo{author}{\bibfnamefont{C.}~\bibnamefont{Chappert}},
  \bibinfo{author}{\bibfnamefont{J.~A.} \bibnamefont{Katine}},
  \bibinfo{author}{\bibfnamefont{M.~J.} \bibnamefont{Carey}}, \bibnamefont{and}
  \bibinfo{author}{\bibfnamefont{K.}~\bibnamefont{Ito}},
  \bibinfo{journal}{Applied Physics Letters} \textbf{\bibinfo{volume}{88}},
  \bibinfo{eid}{192507} (\bibinfo{year}{2006}).

\bibitem[{\citenamefont{Petit et~al.}(2007)\citenamefont{Petit, Baraduc,
  Thirion, Ebels, Liu, Li, Wang, and Dieny}}]{Petit2007}
\bibinfo{author}{\bibfnamefont{S.}~\bibnamefont{Petit}},
  \bibinfo{author}{\bibfnamefont{C.}~\bibnamefont{Baraduc}},
  \bibinfo{author}{\bibfnamefont{C.}~\bibnamefont{Thirion}},
  \bibinfo{author}{\bibfnamefont{U.}~\bibnamefont{Ebels}},
  \bibinfo{author}{\bibfnamefont{Y.}~\bibnamefont{Liu}},
  \bibinfo{author}{\bibfnamefont{M.}~\bibnamefont{Li}},
  \bibinfo{author}{\bibfnamefont{P.}~\bibnamefont{Wang}}, \bibnamefont{and}
  \bibinfo{author}{\bibfnamefont{B.}~\bibnamefont{Dieny}},
  \bibinfo{journal}{Phys. Rev. Lett.} \textbf{\bibinfo{volume}{98}},
  \bibinfo{pages}{077203} (\bibinfo{year}{2007}).

\bibitem[{\citenamefont{Houssameddine et~al.}(2008)\citenamefont{Houssameddine,
  Florez, Katine, Michel, Ebels, Mauri, Ozatay, Delaet, Viala, Folks
  et~al.}}]{HouseSTT}
\bibinfo{author}{\bibfnamefont{D.}~\bibnamefont{Houssameddine}},
  \bibinfo{author}{\bibfnamefont{S.~H.} \bibnamefont{Florez}},
  \bibinfo{author}{\bibfnamefont{J.~A.} \bibnamefont{Katine}},
  \bibinfo{author}{\bibfnamefont{J.-P.} \bibnamefont{Michel}},
  \bibinfo{author}{\bibfnamefont{U.}~\bibnamefont{Ebels}},
  \bibinfo{author}{\bibfnamefont{D.}~\bibnamefont{Mauri}},
  \bibinfo{author}{\bibfnamefont{O.}~\bibnamefont{Ozatay}},
  \bibinfo{author}{\bibfnamefont{B.}~\bibnamefont{Delaet}},
  \bibinfo{author}{\bibfnamefont{B.}~\bibnamefont{Viala}},
  \bibinfo{author}{\bibfnamefont{L.}~\bibnamefont{Folks}},
  \bibnamefont{et~al.}, \bibinfo{journal}{Applied Physics Letters}
  \textbf{\bibinfo{volume}{93}}, \bibinfo{eid}{022505} (\bibinfo{year}{2008}).

\bibitem[{\citenamefont{Houssameddine et~al.}(2009)\citenamefont{Houssameddine,
  Ebels, Dieny, Garello, Michel, Delaet, Viala, Cyrille, Katine, and
  Mauri}}]{HousePRL}
\bibinfo{author}{\bibfnamefont{D.}~\bibnamefont{Houssameddine}},
  \bibinfo{author}{\bibfnamefont{U.}~\bibnamefont{Ebels}},
  \bibinfo{author}{\bibfnamefont{B.}~\bibnamefont{Dieny}},
  \bibinfo{author}{\bibfnamefont{K.}~\bibnamefont{Garello}},
  \bibinfo{author}{\bibfnamefont{J.-P.} \bibnamefont{Michel}},
  \bibinfo{author}{\bibfnamefont{B.}~\bibnamefont{Delaet}},
  \bibinfo{author}{\bibfnamefont{B.}~\bibnamefont{Viala}},
  \bibinfo{author}{\bibfnamefont{M.-C.} \bibnamefont{Cyrille}},
  \bibinfo{author}{\bibfnamefont{J.~A.} \bibnamefont{Katine}},
  \bibnamefont{and} \bibinfo{author}{\bibfnamefont{D.}~\bibnamefont{Mauri}},
  \bibinfo{journal}{Phys. Rev. Lett.} \textbf{\bibinfo{volume}{102}},
  \bibinfo{pages}{257202} (\bibinfo{year}{2009}).

\bibitem[{\citenamefont{Devolder et~al.}(2009)\citenamefont{Devolder,
  Bianchini, Kim, Crozat, Chappert, Cornelissen, Op~de Beeck, and
  Lagae}}]{Devolder2009}
\bibinfo{author}{\bibfnamefont{T.}~\bibnamefont{Devolder}},
  \bibinfo{author}{\bibfnamefont{L.}~\bibnamefont{Bianchini}},
  \bibinfo{author}{\bibfnamefont{J.-V.} \bibnamefont{Kim}},
  \bibinfo{author}{\bibfnamefont{P.}~\bibnamefont{Crozat}},
  \bibinfo{author}{\bibfnamefont{C.}~\bibnamefont{Chappert}},
  \bibinfo{author}{\bibfnamefont{S.}~\bibnamefont{Cornelissen}},
  \bibinfo{author}{\bibfnamefont{M.}~\bibnamefont{Op~de Beeck}},
  \bibnamefont{and} \bibinfo{author}{\bibfnamefont{L.}~\bibnamefont{Lagae}},
  \bibinfo{journal}{Journal of Applied Physics} \textbf{\bibinfo{volume}{106}},
  \bibinfo{eid}{103921} (\bibinfo{year}{2009}).

\bibitem[{\citenamefont{Thadani et~al.}(2008)\citenamefont{Thadani, Finocchio,
  Li, Ozatay, Sankey, Krivorotov, Cui, Buhrman, and Ralph}}]{Thadani2008}
\bibinfo{author}{\bibfnamefont{K.~V.} \bibnamefont{Thadani}},
  \bibinfo{author}{\bibfnamefont{G.}~\bibnamefont{Finocchio}},
  \bibinfo{author}{\bibfnamefont{Z.-P.} \bibnamefont{Li}},
  \bibinfo{author}{\bibfnamefont{O.}~\bibnamefont{Ozatay}},
  \bibinfo{author}{\bibfnamefont{J.~C.} \bibnamefont{Sankey}},
  \bibinfo{author}{\bibfnamefont{I.~N.} \bibnamefont{Krivorotov}},
  \bibinfo{author}{\bibfnamefont{Y.-T.} \bibnamefont{Cui}},
  \bibinfo{author}{\bibfnamefont{R.~A.} \bibnamefont{Buhrman}},
  \bibnamefont{and} \bibinfo{author}{\bibfnamefont{D.~C.} \bibnamefont{Ralph}},
  \bibinfo{journal}{Phys. Rev. B} \textbf{\bibinfo{volume}{78}},
  \bibinfo{pages}{024409} (\bibinfo{year}{2008}).

\bibitem[{\citenamefont{Cascales et~al.}(2013)\citenamefont{Cascales, Herranz,
  Sambricio, Ebels, Katine, and Aliev}}]{CascalesAPL2013}
\bibinfo{author}{\bibfnamefont{J.~P.} \bibnamefont{Cascales}},
  \bibinfo{author}{\bibfnamefont{D.}~\bibnamefont{Herranz}},
  \bibinfo{author}{\bibfnamefont{J.~L.} \bibnamefont{Sambricio}},
  \bibinfo{author}{\bibfnamefont{U.}~\bibnamefont{Ebels}},
  \bibinfo{author}{\bibfnamefont{J.~A.} \bibnamefont{Katine}},
  \bibnamefont{and} \bibinfo{author}{\bibfnamefont{F.~G.} \bibnamefont{Aliev}},
  \bibinfo{journal}{Applied Physics Letters} \textbf{\bibinfo{volume}{102}},
  \bibinfo{eid}{092404} (\bibinfo{year}{2013}).

\bibitem[{\citenamefont{Cascales}(2015)}]{tesisjp}
\bibinfo{author}{\bibfnamefont{J.~P.} \bibnamefont{Cascales}}, Ph.D. thesis,
  \bibinfo{school}{Universidad Aut\'onoma de Madrid, Spain}
  (\bibinfo{year}{2015}).

\bibitem[{\citenamefont{Houssameddine}(2009)}]{ThesisHousse}
\bibinfo{author}{\bibfnamefont{D.}~\bibnamefont{Houssameddine}},
  \bibinfo{type}{Theses}, \bibinfo{school}{{Universit{\'e} Joseph-Fourier -
  Grenoble I}} (\bibinfo{year}{2009}).

\bibitem[{\citenamefont{Yuasa et~al.}(2004)\citenamefont{Yuasa, Nagahama,
  Fukushima, Suzuki, and Ando}}]{Yuasa2004}
\bibinfo{author}{\bibfnamefont{S.}~\bibnamefont{Yuasa}},
  \bibinfo{author}{\bibfnamefont{T.}~\bibnamefont{Nagahama}},
  \bibinfo{author}{\bibfnamefont{A.}~\bibnamefont{Fukushima}},
  \bibinfo{author}{\bibfnamefont{Y.}~\bibnamefont{Suzuki}}, \bibnamefont{and}
  \bibinfo{author}{\bibfnamefont{K.}~\bibnamefont{Ando}},
  \bibinfo{journal}{Nature Materials} \textbf{\bibinfo{volume}{3}},
  \bibinfo{pages}{868} (\bibinfo{year}{2004}).

\bibitem[{\citenamefont{Alvarez-H{\'e}rault}(2010)}]{ThesisAlvarez}
\bibinfo{author}{\bibfnamefont{J.}~\bibnamefont{Alvarez-H{\'e}rault}},
  \bibinfo{type}{Theses}, \bibinfo{school}{{Universit{\'e} de Grenoble}}
  (\bibinfo{year}{2010}).

\bibitem[{\citenamefont{Helmer et~al.}(2010)\citenamefont{Helmer, Cornelissen,
  Devolder, Kim, van Roy, Lagae, and Chappert}}]{Helmer2010}
\bibinfo{author}{\bibfnamefont{A.}~\bibnamefont{Helmer}},
  \bibinfo{author}{\bibfnamefont{S.}~\bibnamefont{Cornelissen}},
  \bibinfo{author}{\bibfnamefont{T.}~\bibnamefont{Devolder}},
  \bibinfo{author}{\bibfnamefont{J.-V.} \bibnamefont{Kim}},
  \bibinfo{author}{\bibfnamefont{W.}~\bibnamefont{van Roy}},
  \bibinfo{author}{\bibfnamefont{L.}~\bibnamefont{Lagae}}, \bibnamefont{and}
  \bibinfo{author}{\bibfnamefont{C.}~\bibnamefont{Chappert}},
  \bibinfo{journal}{Phys. Rev. B} \textbf{\bibinfo{volume}{81}},
  \bibinfo{pages}{094416} (\bibinfo{year}{2010}).

\bibitem[{\citenamefont{Eklund et~al.}(2014)\citenamefont{Eklund, Bonetti,
  Sani, Majid~Mohseni, Persson, et~al.}}]{Eklund2014}
\bibinfo{author}{\bibfnamefont{A.}~\bibnamefont{Eklund}},
  \bibinfo{author}{\bibfnamefont{S.}~\bibnamefont{Bonetti}},
  \bibinfo{author}{\bibfnamefont{S.~R.} \bibnamefont{Sani}},
  \bibinfo{author}{\bibfnamefont{S.}~\bibnamefont{Majid~Mohseni}},
  \bibinfo{author}{\bibfnamefont{J.}~\bibnamefont{Persson}},
  \bibnamefont{et~al.}, \bibinfo{journal}{Applied Physics Letters}
  \textbf{\bibinfo{volume}{104}}, \bibinfo{eid}{092405} (\bibinfo{year}{2014}).

\bibitem[{\citenamefont{Sharma et~al.}(2014)\citenamefont{Sharma, Dürrenfeld,
  Iacocca, Heinonen, Åkerman, and Muduli}}]{Sharma2014}
\bibinfo{author}{\bibfnamefont{R.}~\bibnamefont{Sharma}},
  \bibinfo{author}{\bibfnamefont{P.}~\bibnamefont{Dürrenfeld}},
  \bibinfo{author}{\bibfnamefont{E.}~\bibnamefont{Iacocca}},
  \bibinfo{author}{\bibfnamefont{O.~G.} \bibnamefont{Heinonen}},
  \bibinfo{author}{\bibfnamefont{J.}~\bibnamefont{Åkerman}}, \bibnamefont{and}
  \bibinfo{author}{\bibfnamefont{P.~K.} \bibnamefont{Muduli}},
  \bibinfo{journal}{Applied Physics Letters} \textbf{\bibinfo{volume}{105}},
  \bibinfo{pages}{132404} (\bibinfo{year}{2014}).

\bibitem[{\citenamefont{Gokce et~al.}(2006)\citenamefont{Gokce, Nowak, Yang,
  and Parkin}}]{Gokce2006}
\bibinfo{author}{\bibfnamefont{A.}~\bibnamefont{Gokce}},
  \bibinfo{author}{\bibfnamefont{E.~R.} \bibnamefont{Nowak}},
  \bibinfo{author}{\bibfnamefont{S.~H.} \bibnamefont{Yang}}, \bibnamefont{and}
  \bibinfo{author}{\bibfnamefont{S.~S.~P.} \bibnamefont{Parkin}},
  \bibinfo{journal}{Journal of Applied Physics} \textbf{\bibinfo{volume}{99}},
  \bibinfo{pages}{08} (\bibinfo{year}{2006}).

\bibitem[{\citenamefont{Aliev et~al.}(2007)\citenamefont{Aliev, Guerrero,
  Herranz, Villar, Greullet, Tiusan, and Hehn}}]{Aliev2007}
\bibinfo{author}{\bibfnamefont{F.~G.} \bibnamefont{Aliev}},
  \bibinfo{author}{\bibfnamefont{R.}~\bibnamefont{Guerrero}},
  \bibinfo{author}{\bibfnamefont{D.}~\bibnamefont{Herranz}},
  \bibinfo{author}{\bibfnamefont{R.}~\bibnamefont{Villar}},
  \bibinfo{author}{\bibfnamefont{F.}~\bibnamefont{Greullet}},
  \bibinfo{author}{\bibfnamefont{C.}~\bibnamefont{Tiusan}}, \bibnamefont{and}
  \bibinfo{author}{\bibfnamefont{M.}~\bibnamefont{Hehn}},
  \bibinfo{journal}{Applied Physics Letters} \textbf{\bibinfo{volume}{91}},
  \bibinfo{eid}{232504} (\bibinfo{year}{2007}).

\bibitem[{\citenamefont{Almeida et~al.}(2008)\citenamefont{Almeida, Wisniowski,
  and Freitas}}]{Almeida2008}
\bibinfo{author}{\bibfnamefont{J.~M.} \bibnamefont{Almeida}},
  \bibinfo{author}{\bibfnamefont{P.}~\bibnamefont{Wisniowski}},
  \bibnamefont{and} \bibinfo{author}{\bibfnamefont{P.}~\bibnamefont{Freitas}},
  \bibinfo{journal}{Magnetics, IEEE Transactions on}
  \textbf{\bibinfo{volume}{44}}, \bibinfo{pages}{2569} (\bibinfo{year}{2008}).

\end{thebibliography}
\end{document}